\documentclass[12pt,preprint]{aastex}
\def\change#1{{\it #1}}
\DeclareTextSymbol{\degre}{OT1}{23}

\slugcomment{}

\shorttitle{Origin of heavy noble gases in terrestrial planets}
 
\shortauthors{Mousis et al.}

\begin{document}

\title{Impact regimes and post-formation sequestration processes: implications for the origin of heavy noble gases in terrestrial planets}

\author{
Olivier~Mousis\altaffilmark{1},
Jonathan I.~Lunine\altaffilmark{2},
Jean-Marc~Petit\altaffilmark{1},
Sylvain~Picaud\altaffilmark{1},
Bernard~Schmitt\altaffilmark{3},
Didier~Marquer\altaffilmark{4},
Jonathan~Horner\altaffilmark{5,6}
and Caroline~Thomas\altaffilmark{1}}

\altaffiltext{1}{Universit{\'e} de Franche-Comt{\'e}, Institut UTINAM, CNRS/INSU, UMR 6213, 25030 Besan\c{c}on Cedex, France}

\email{olivier.mousis@obs-besancon.fr}

\altaffiltext{2}{Dipartimento di Fisica, Universit{\`a} degli Studi di Roma ``Tor Vergata'', Rome, Italy}

\altaffiltext{3}{Universit{\'e} Joseph Fourier, Laboratoire de Plan{\'e}tologie de Grenoble, CNRS/INSU, UMR 5109, Observatoire de Sciences de l'Univers de Grenoble, France}

\altaffiltext{4}{Universit{\'e} de Franche-Comt{\'e}, Chrono-Environnement, CNRS/INSU, UMR 6249, 25030 Besan\c con Cedex, France}

\altaffiltext{5}{Department of Physics, Science Laboratories, University of Durham, South Road, Durham, DH1 3LE, UK}

\altaffiltext{6}{Astronomy Group, Physics \& Astronomy, The Open University, Milton Keynes, MK7 6AA, UK}

\begin{abstract}
{The difference between the {measured atmospheric} abundances of neon, argon, krypton and xenon {for} Venus, {the} Earth and Mars is striking. Because these abundances {drop by at least two orders of magnitude as one moves} outward from Venus to Mars, the study of the origin of this discrepancy is a key issue {that must be explained if we are to fully understand} the different delivery mechanisms of the volatiles accreted by the terrestrial planets. In this work, we aim {to investigate whether} it is possible to quantitatively explain the {variation} of the heavy noble gas abundances measured {on} Venus, {the} Earth and Mars, assuming that cometary bombardment was the main delivery mechanism of these noble gases to the terrestrial planets. To do so, we use recent dynamical simulations that {allow the study of} the impact fluxes of comets {upon} the terrestrial planets during the course of their formation and evolution. Assuming that the mass of noble gases delivered by comets is proportional to {rate at which they collide with the terrestrial planets}, we show that the krypton and xenon abundances in Venus and {the} Earth can be explained in a manner consistent with the hypothesis of cometary bombardment. In order to explain the krypton and xenon abundance differences between {the} Earth and Mars, we need to invoke the presence of large amounts of CO$_2$-dominated clathrates in the Martian soil that would have efficiently sequestered these noble gases.
Two different scenarios based on our model can be used to also explain the differences {between} the neon and argon abundances {of} the terrestrial planets. In the first scenario, cometary bombardment of these planets would have occurred at epochs contemporary {with} the existence of their primary atmospheres. Comets would have been the carriers of argon, krypton and xenon while neon would have been gravitationally captured by the terrestrial planets.  In the second scenario, {we consider impacting comets that contained significantly smaller amounts of argon}, an idea supported by predictions of noble gas abundances in these bodies, provided {that} they formed from clathrates in the solar nebula. In this scenario, neon and argon would have been supplied to the terrestrial planets via the gravitational capture of their primary atmospheres whereas {the bulk of their} krypton and xenon would have been delivered by comets.}

\end{abstract}


\section{Introduction}
\label{Intro}

The difference between the measured atmospheric abundances of non-radiogenic noble gases in Venus, Earth and Mars is striking. {It is well known that these abundances decline dramatically as one moves outward from Venus to Mars within the inner Solar system, with these two planets differing in abundance by up to two orders of magnitude (see Fig. 1 and Table \ref{ratio}). Therefore, understanding this variation is a key issue in understanding how the primordial atmospheres of the terrestrial planets evolved to their current composition, and requires us to study} the different delivery mechanisms of the volatiles accreted by these planets (Pepin 1991, 2006; Owen et al. 1992, 1995; Dauphas 2003; Marty and Meibom 2007). Another striking feature of the atmospheric noble gas abundances in terrestrial planets is that they {all exhibit significant depletion} relative to solar composition (see Fig. 1 and Pepin 1991, 2006). This depletion depends {, at a first approximation, on the mass of the element in question, with} the light gases being more depleted and isotopically fractionated than the heavy ones. Moreover, the nature of the main source of noble gases remains controversial. Indeed, only a relatively small fraction was supplied to the atmospheres of terrestrial planets via the outgassing of their mantles (Marty \& Meibom 2007). On the other hand, it has been proposed that noble gases could have been acquired by {the} terrestrial planets {through} the gravitational capture of their primary atmospheres (Pepin 1991, 2006). These {gases would have initially been captured in approximately solar-like abundances, then experienced ongoing fractionation as a result of} gravitational escape, driven on Earth by a giant Moon-forming impact, on Mars by sputtering at high altitudes {(and also, potentially, the putative giant impact which formed the observed Martian crustal dichotomy (Andrews-Hanna et al. 2008)). In addition, the fractionation process will also have been driven by the absorption of intense ultraviolet radiation from the young Sun (Pepin 2006) on each of the terrestrial planets}. Other scenarios suggest that bombardment by icy planetesimals or comets could be the main delivery mechanism of argon, krypton and xenon to the terrestrial planets (Owen et al. 1992; Owen and Bar-Nun 1995). In these scenarios, the idea {that} an asteroidal bombardment could be the main contributor of noble gases to the terrestrial planets is not favored because the trend described by the chondritic noble gas abundances as a function of their atomic mass does not reflect those observed in Venus, Earth and Mars (Owen and Bar-Nun 1995). In order to account for the differences {between the} noble gas abundances observed today among these planets, Owen et al. (1992) argued that giant impacts would have eroded their atmospheres without inducing any fractionation. In their model, gravitational escape would have played a role restricted to the fractionation of atmospheric neon. The idea that two {separate} sources {(i.e. fractionated nebular gases and accreted cometary volatiles)} contributed to the current noble gas budget on Earth has also been developed by Dauphas (2003). In this case, hydrothermal escape would have fractionated the heavy noble gases acquired from the solar nebula and the resulting transient atmosphere would have mixed with cometary material in order to reproduce the Earth's noble gas abundances.

In this work, we {investigate whether} it is possible to quantitatively explain the differences {between} the heavy noble gas abundances measured {in the atmospheres of Venus, the Earth} and Mars, assuming that cometary bombardment was the main delivery mechanism of these noble gases to the terrestrial planets. To do so, we use the recent dynamical simulations performed by Horner et al. (2009) that allowed {the calculation of} the impact fluxes of small bodies suffered by the terrestrial planets during the course of their formation and evolution. Assuming that the mass of noble gases delivered by comets is proportional to the {rate at which they impact upon the terrestrial planets}, we show that the krypton and xenon abundances in Venus and Earth can be explained in a manner consistent with the hypothesis of cometary bombardment. In order to explain the {differences between the} krypton and xenon abundances {of the} Earth and Mars, we need to invoke the presence of large amounts of CO$_2$-dominated clathrates in the Martian soil that would have efficiently sequestered these noble gases. On the other hand, our calculations show that it is impossible {for} the variation of the argon abundance between the terrestrial planets {to be explained if the sole source of volatile material is cometary in nature}. In order to account for this variation, a loss of argon due to gravitational escape needs to be invoked {from} the primary atmospheres of the terrestrial planets.

\section{The impact regimes of the terrestrial planets}
\label{Impact}

Recent $n$-body simulations performed by Horner et al. (2009) {studied the impact rates experienced by the terrestrial planets as a result of diverse populations of potential impactors.} They considered a wide range of plausible planetary formation scenarios for the terrestrial planets, and found that the different impact regimes {experienced by} Venus, {the} Earth and Mars {could have resulted in significant differences between their individual hydration states over the course of their formation and evolution}. These authors created different test populations of massless particles representing asteroidal and cometary material {, and followed their dynamical evolution for a period of 10 Myr under the gravitational influence of the planets} (see Horner et al. 2009 for further details). For the sake of comparison, Table \ref{ratio1} shows the results of two {of the} simulations conducted by Horner et al. (2009), {investigating the impact flux of cometary bodies on the terrestrial planets}. Simulation 1 describes the evolution of a population {of 100,000} objects ($N_{com-init}$) analogous to the Centaurs in the modern outer solar system (5$<$$q$$<$10 AU, 5$<$$Q$$<$30 AU and 0$\degre<i<$30$\degre$; see e.g. Levison \& Duncan 1997; Horner et al. 2003,  2004; Bailey \& Malhotra 2009 and Jewitt 2009 for a variety of papers on the nature of the Centaurs). Simulation 2 describes the simultaneous evolution of two different populations of 50,000 comets each. The first group of objects ($N_{JNS}$) were created with semi-major axes between 6 and 10 AU, eccentricities less than 0.2, and inclinations less than 10$\degre$. The second group ($N_{SNN}$) were created with the same spread of eccentricities and inclinations, but with semi-major axes between 10 and 30 AU. Table \ref{ratio1} shows that the results of both simulations are very similar. {On average, the Earth received a flux of impacting comets which is $\sim$3.4 times higher than that experienced by Mars. On the other hand, Venus experiences an impact flux of comets that is only $\sim$0.8 times that for the Earth.}

{In this work, we assume that the mass of noble gases delivered by comets to the terrestrial planets is proportional to the rate at which they impacted upon them}. It is then possible to derive $X_p / X_{p'}$ {(the ratio of the noble gas abundances ({as a fraction of the total mass} of the planet) between planets $p$ and $p'$)} from the ratio {of the number of comets impacting upon those two planets $N_p / N_{p'}$, (see Table  \ref{ratio2}) through} the following relation:

\begin{equation}
\frac{X_{p}}{X_{p'}} = \frac{N_{p}}{N_{p'}} \frac{M_{p'}}{M_p},
\end{equation}

\noindent where $M_p$ and $M_{p'}$ are the masses of planets $p$ and $p'$, respectively. Table \ref{ratio2} displays the values of $X_p / X_{p'}$ for the terrestrial planets and shows that the {amount of noble gases delivered to the Earth and Venus, as a fraction of their total mass, is almost identical.} Assuming that comets were the only {source} of the heavier noble gases to the terrestrial planets, this result {falls within} the range of values {obtained through} observation {(Table \ref{ratio}) of} the $^{84}$Kr abundance in Venus and the Earth. 

{\bf Because of the large uncertainties involved in the measurement of the abundance of noble gases in the atmosphere of Venus, we also consider that the ratio of the $^{130}$Xe abundances of the two planets is {in agreement with} our model. Indeed, the $^{130}$Xe abundance used in Table \ref{NG} and derived from Pepin (1991) corresponds to the value determined by Donahue (1986) from the data collected by the U.S. Pioneer Venus probe. On the other hand, the same data and others collected by Soviet {\bf spacecraft}  have been comprehensively reviewed by von Zahn et al. (1983) who suggested that xenon might not have been detected at all. Moreover, other determinations by Donahue et al. (1981){\bf, still from the same Pioneer Venus data set,} also suggest that the measured $^{130}$Xe abundance is an upper limit and could be as low as zero. In other words, all these studies suggest that the Venusian data are not in a state to permit a robust determination of the Xe isotope abundances.}

On the other hand, our model cannot reproduce the high ratio (up to two orders of magnitude) between the observed $^{36}$Ar abundances in Venus and the Earth. Moreover, Table \ref{ratio2} shows that the average noble gas abundance on {the} Earth should be $\sim$ 0.37 times the Martian noble gas abundances if these volatiles were {solely} delivered by comets. This result {differs significantly from that} inferred from measurements of noble gas abundances, which {are observed to be approximately} two orders of magnitude larger for the Earth compared to Mars. From these considerations, it appears that the different impact regimes undergone by the terrestrial planets are not sufficient to explain all the variations between their noble gas abundances.

In the next Section, we discuss the potential influence that multiple guest clathrates located in the Martian subsurface could have on its atmospheric noble gas budget.

\section{\change{The trapping of noble gases by clathrates in the Martian soil}}

Theories supporting the presence of clathrates on Mars initially focused on CO$_2$ as the sequestered gas, but have been extended to the investigation of the trapping of other gases. It was shown in 1970 that CO$_2$ clathrate was thermodynamically stable at the Martian poles (Miller and Smythe 1970). Since {then}, many studies have postulated the existence of large CO$_2$ clathrate deposits {both} at the poles {and in the subsurface at lower latitudes,} and discussed their possible influence on the Martian climate or atmospheric composition (Milton, 1974; Dobrovolskis and Ingersoll 1975; Jakosky et al. 1995; Musselwhite and Lunine 1995; Kargel and Lunine 1998; Komatsu et al. 2000; Max and Clifford 2001; Longhi 2006; Prieto-Ballesteros et al. 2006; Chastain and Chevrier 2007; Thomas et al. 2009; Swindle et al. 2009).

The {presence} of CO$_2$-dominated clathrates in the {perennial} Martian polar caps or in the permafrost is important for our study because it could imply that the $^{36}$Ar, $^{84}$Kr and $^{130}$Xe abundances measured in the planet's atmosphere are not representative of its global noble gas budget. Indeed, depending on the amount of existing clathrates, the {volume} of noble gases trapped in these crystalline structures could be much larger than those measured in the atmosphere. This was shown by Musselwhite and Lunine (1995) and Swindle et al. (2009) who exploited a statistical mechanical model (van der Waals \& Platteeuw 1959; Lunine and Stevenson 1985) to determine the composition of CO$_2$-dominated clathrates formed on the Martian surface. In this model, the relative abundance $f_G$ of a guest species $G$ in a clathrate (of structure I because CO$_2$ is the primary guest species) is defined as the ratio of the average number of guest molecules of species $G$ in the clathrate over the average total number of incorporated molecules, as:

\begin{equation}
\label{abondance} f_G=\frac{b_L y_{G,L}+b_S y_{G,S}}{b_L
\sum_J{y_{J,L}}+b_S \sum_J{y_{J,S}}},
\end{equation}

\noindent where the sums in the denominator run over all species present in the system, and $b_S$ and $b_L$ are the number of small and large cages per unit cell, respectively. The occupancy fractions $y_G$ of the guest species $G$ for a given type of cage and for a given type of clathrate are determined from the Langmuir constants which are related to the strength of the interaction between each guest species and each type of cage (van der Waals \& Platteeuw 1959; Lunine and Stevenson 1985). In a first approximation, this cage is assumed to be spherical and the corresponding interactions are represented by a spherically averaged Kihara potential, the integration of which within the cage giving the Langmuir constants. The calculation via Eq. \ref{abondance} of the relative abundance of a guest species trapped in clathrate depend both on the structural characteristics of the considered clathrate and on the accuracy of the corresponding interactions between the trapped molecule and the water cage. As a consequence, the accuracy of the calculations strongly depends on the choice of the parameters of the Kihara potential (Papadimitriou et al. 2007; Thomas et al. 2008). To determine the Kihara parameters, Lunine and Stevenson (1985) fitted experimental dissociation pressures directly, while Parrish and Prauznitz (1972a,b) used fits to the chemical potential difference between water ice (or liquid) and clathrate hydrate determined by experimental data. Extrapolated down to low temperatures, small differences in the fits at higher temperature would become magnified between the two approaches. Furthermore, while Lunine and Stevenson (1985) assumed a constant value with temperature for the chemical potential difference between water ice and the empty clathrate hydrate cage, Parrish and Prausnitz derived a temperature dependence from the experimental data.  Although this would not affect the predictions of clathrate formation in the pure noble gases, CO$_2$ has such a large, rod-like core that as a primary guest species it could lead to large differences in noble gas trapping efficiencies depending on the two approaches.  Indeed there are significant differences in trapping efficiency between the Kihara parameters derived by the two sets of authors. For example, from a set of Kihara parameters determined by Lunine and Stevenson (1985), Musselwhite and Lunine (1995) found that the mole fractions of Xe, Kr and Ar in CO$_2$-dominated clathrates are $\sim$ 700, 6 and 0.1 times those measured in the Martian atmosphere, for a soil temperature of 150 K and a CO$_2$ gas pressure of 0.007 bar, i.e. the atmospheric pressure at the ground level. On the other hand, from another set of Kihara parameters self-consistently determined by Parrish and Prausnitz (1972a,b) on some experimentally measured dissociation pressures, Swindle et al. (2009) found that the mole fractions of Xe, Kr and Ar in CO$_2$-dominated clathrates are $\sim$ 34, 0.6 and 0.02 times those measured in the Martian atmosphere, in the same clathration conditions. Despite the discrepancies, these calculations confirm that at least large amounts of xenon and krypton can be trapped in the CO$_2$-dominated Martian clathrates.

The amount of noble gases potentially sequestered in the Martian soil will depend on the postulated mass of existing clathrates. Using Mariner 9 and Viking Orbiter observations, Jakosky et al. (1995) assumed a total volume of the north polar layered deposits (NPLD) of $\sim 2 \times 10^{6}$ km$^3$ and inferred that up to the equivalent of $\sim$0.2 bar of CO$_2$ could be contained in the NPLD if it consists primarily of clathrates. However, recent observations by the sounding radar SHARAD on the Mars Reconnaissance Orbiter have shown that the NPLD should be somewhat smaller, with an estimated volume of $\sim 1.1 \times 10^{6}$ km$^3$ (Grima et al. 2009). On the other hand, observations of the south-polar layered deposits (SPLD) by the sounding radar MARSIS aboard the Mars Express orbiter indicated the plausible presence of $\sim 1.6 \times 10^{6}$ km$^3$ of ice (Plaut et al. 2007). From these recent estimates and following the approach of Jakosky et al. (1995), if all the polar deposits consist essentially of clathrates, then the equivalent pressure of CO$_2$ trapped in clathrates could reach $\sim$0.27 bar, provided that all the clathrates cages are occupied. Note that these estimates do not take into account the possible existence of clathrates in the Martian permafrost whose thickness is expected to vary between 3--5 km at the equator to 8--13 km at the poles (Max and Clifford 2000, 2004; Chastain and Chevrier 2007). Assuming that only 10\% of this layer is made of ice would add between $\sim$30 and 50 $\times$ 10$^{6}$ km$^3$ of ice and thus more than 3 bar of CO$_2$ if  all this permafrost is made of CO$_2$ clathrate. However, {as of yet,} there is no observation of such a deep and extended permafrost layer. Recent MARSIS analysis demonstrated that an ice-rich permafrost layer exists close to the surface with a depth of at least 50 m at all latitudes poleward of $\pm$ 50$\degre$ in both hemispheres, adding $\sim 1 \times 10^{6}$ km$^3$ to the known ice budget (Mouginot et al. 2009, submitted). However the dielectric constants of water ice and clathrates are too similar to allow to them to be easily distinguished at these radar wavelengths. If we assume this ice layer is also clathrates, this adds an atmospheric equivalent of 0.1 bar of CO$_2$. The total CO$_2$ content of all known ice on Mars could be up to 0.37 bar. However, because the efficiency of clathration is never total, we will assume here that the trapping of 0.3 bar of CO$_2$ in Martian clathrates is plausible. It is an upper limit for the known Martian ice but also a lower limit given all the still-unknown but probably deep underground ice reservoirs.

This value implies that, if the mixing ratios of guest species trapped in clathrates were similar to the atmospheric ones, about 0.3/0.007 $\sim$ 43 times more noble gases would be present in the Martian subsurface than in the atmosphere. Taking into account the fractionation of the noble gas abundances induced by their clathration, the amount of noble gases stored in the Martian subsurface could be much larger. In this case, if one considers the Kihara parameters derived by Lunine and Stevenson (1985), the equivalent abundances of Xe, Kr and Ar permanently stored in the Martian subsurface could be up to $\sim$ 30,000, 260 and 4 times higher than their observed atmospheric abundances for 0.3 bar of CO$_2$ trapped in clathrates. 

{Calculations detailing the impact regimes of the terrestrial planets} predict that, if comets were the main suppliers of noble gases to the terrestrial planets, then the average noble gas abundance (scaled to the mass of the planet) on Earth {should be approximately} 0.37 times the noble gas abundance on Mars. In this case, the equivalent of $\sim$ 216 to 307 times the Martian $^{84}$Kr atmospheric abundance is required to be trapped in clathrates in order to match the observed Earth/Mars $\sim$80--113 range of $^{84}$Kr abundance ratios. This value remains below the maximum predicted from the assumption that 0.3 bar of CO$_2$ is stored in the Martian clathrates and that all the clathrate cages are filled by guest molecules. Similarly, if the equivalent of $\sim$150-230 times the Martian atmospheric abundance of  $^{130}$Xe is trapped in clathrates, a range of values well below the maximum predicted, then the resulting terrestrial  $^{130}$Xe abundance is in agreement with the observed Earth/Mars $\sim$55--85 range of $^{130}$Xe abundance ratios. However, the Earth/Mars $^{36}$Ar observed range of ratios of 127--215 cannot be explained by our model because the trapping of this species is very poor in Martian clathrates, at least $\sim$ 86 times less than that would be required to reduce the Martian atmospheric argon to its current level.

In order to match the observations, the amounts of krypton and xenon trapped in CO$_2$-dominated clathrates must be enhanced by a similar factor compared to their atmospheric abundances. This conclusion could appear to contradict the present results that show that the maximum trapping efficiencies of krypton and xenon calculated here are very different. However, it should be noted that these differences could be strongly reduced by considering the variation of the cage size with temperature. Indeed, it has been recently shown that a small contraction of the cage size strongly increases the trapping efficiency of Kr without changing noticeably the one of Xe, at least for CH$_4$-rich clathrates on Titan (see Fig. 3 of Thomas et al. 2008). Moreover, the present calculations have been done using Kihara parameters given by the same authors for all species. However, each set of Kihara parameters describing the interaction potential of a given molecule with the clathrate cage is in fact individually determined. With our current state of knowledge it is still difficult to determine which published parameter set is the most appropriate and it is possible to select Kihara parameters among those proposed by various authors for krypton and xenon which might yield very similar trapping efficiencies of these noble gases in clathrates.

\section{Discussion and conclusions}

Our model only accounts for the differences between the krypton and xenon abundances in terrestrial planets. However, {in addition,} two different scenarios based on our model can be used to explain the differences {between} the neon and argon abundances of these planets. In the first scenario, cometary bombardment of the terrestrial planets would have occurred at epochs contemporary to the existence of their primary atmospheres. Comets would have been the carriers of argon, krypton and xenon, while neon would have been gravitationally captured by the terrestrial planets (Owen et al. 1992). Only neon and argon would have been fractionated due to gravitational escape, while the abundances of the heavier noble gases would have been poorly affected by {such losses}. In this scenario, the combination of processes such as escape of neon and argon, cometary bombardment at the epochs of existence of primary {planetary} atmospheres and the sequestration of krypton and xenon in the Martian clathrates would then explain the observed noble gas abundances differences between the terrestrial planets. However, this scenario leads to an important chronological issue because the primordial atmospheres of terrestrial planets are expected to exist during the first few million years or so of their lifetime (Pepin 2006), while recent work suggests that heavy noble gases could have been supplied by the late heavy bombardment (Marty and Meibom 2007) which {is thought to have} marked the end of terrestrial planets accretion 3.8 billion years ago (Gomes et al. 2005).

In the second scenario, impacting comets would {be significantly depleted in argon, compared to the other noble gases. This idea is} supported by predictions of the noble gas abundance in these bodies, provided that they formed from clathrates in the solar nebula (Iro et al. 2003). In this model, neon and argon would have been supplied to the terrestrial planets via the gravitational capture of their primary atmospheres whereas {the bulk of their} krypton and xenon would have been delivered by comets. In this case, the cometary bombardment of the terrestrial planets could have occurred {after the formation} of their primary atmospheres because only the neon and argon abundances observed today would have been engendered by the escape-fractionation processes in these atmospheres. Here, we favor this scenario because it is consistent with the hypothesis that the heavier noble gases could have been supplied by the late heavy bombardment to the terrestrial planets.

Whatever the scenario envisaged, our work does not preclude the possibility that a fraction of the heavy noble gases could have been captured by terrestrial planets during the acquisition of their primary atmospheres. In the first scenario, the fraction of krypton and xenon accreted in this way should be low compared to the amount supplied by comets since these noble gases are not expected to have been strongly fractionated by gravitational escape. In the second scenario, the fraction of krypton and xenon captured gravitationally by the terrestrial planets could be large if the escape was efficient. 

{\bf The hypothesis of atmospheric escape is supported by both Mars (SNC meteorites) and Earth, which show substantial fractionation of the xenon isotopes compared to the plausible primitive sources of noble gases, i.e., solar wind (SW--Xe), meteorites (Q--Xe), or the hypothetical U-Xe source\footnote{The U-Xe source is presumably a nebular composition inferred to be present on early Earth but so far not seen directly elsewhere (Pepin 2000).} (Pepin 2006). This fractionation then suggests important losses of Xe and other noble gases from the atmospheres of Earth and Mars. These loss processes might have been more important for Earth and Mars than Venus because the latter planet would have  escaped impacts of the magnitude that formed the {\bf Moon} (Canup \& Asphaug 2001) or created the largest basins on Mars (Andrews-Hanna et al. 2008). In the case of Earth, the noble gas fractionation episode could have been {\bf driven} by impacts (Pepin 1997; Dauphas 2003) or by extreme-ultraviolet radiation (EUV) radiation (Pepin 1991). In the case of Mars, the combination of impacts, hydrodynamic escape due to EUV radiation, planetary degassing and fractionation by sputtering losses (Jakosky et al. 1994; Luhmann et al. 1992; Carr 1999; Chassefi{\`e}re \& Leblanc 2004) might have played an important role in sculpting the {\bf pattern} of the noble gas abundances observed today. On the other hand, EUV from the young Sun that powered Ne-only loss from Earth (Pepin 2006) could have driven escape of Kr and lighter species -- but not Xe -- from the closer Venus (Pepin 1997). Therefore, the present distribution of noble gas abundances in terrestrial planets might result from a combination between the different cometary input fluxes, volatiles sequestration into Martian clathrates and subsequent vigorous atmospheric escapes that occurred during the evolution of these planets.}

Several ongoing space missions could provide {an opportunity} to test the different scenarios proposed in this work. Indeed, the ESA Rosetta spacecraft should provide precise measurements of the noble gas content in Comet 67P/Churyumov-Gerasimenko. These measurements should {allow us to determine whether} argon is depleted in comets (Iro et al. 2003). The NASA Mars Science Laboratory Mission may help evaluate the possibility that large amounts of xenon and krypton are trapped in Martian clathrates (Swindle et al. 2009). Indeed, seasonal variations in the Kr/CO$_2$ and Xe/CO$_2$ ratios should {reveal the extent to which clathrates} participate in the exchange of these noble gases between the {Martian surface and} atmosphere (Swindle et al. 2009).

\acknowledgements
We acknowledge Fran\c cois Forget for fruitful discussions about the evolution of the Martian climate. This work was supported in part by the French Centre National d'Etudes Spatiales, and the program `Incentivazione alla mobilita' di studiosi straineri e italiani residenti all'estero. We acknowledge an anonymous Referee whose comments have helped us to improve our manuscript.

\clearpage

\begin{figure}
\resizebox{\hsize}{!}{\includegraphics[angle=0]{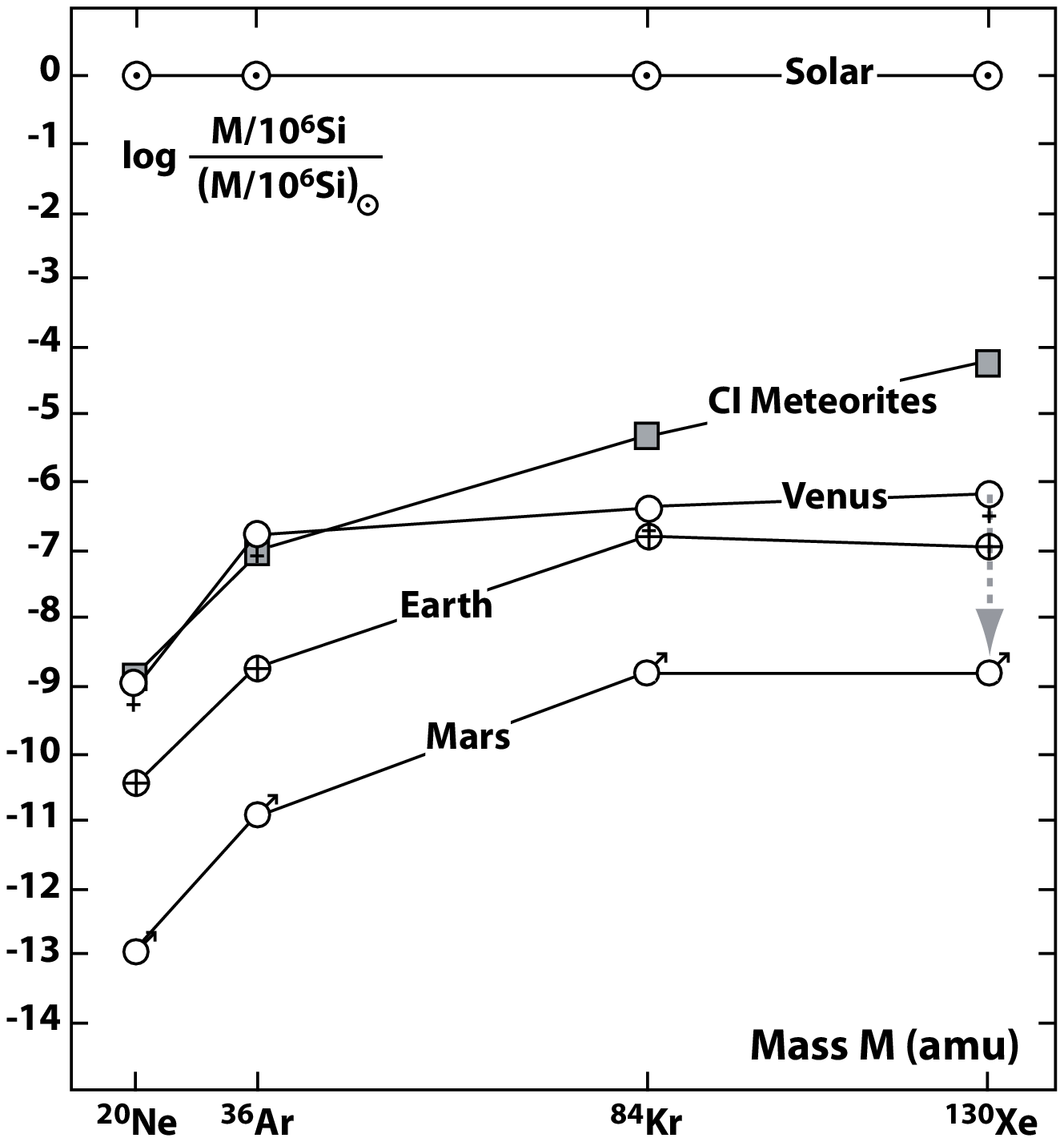}}
\caption{{The measured} abundances of Ne, Ar, Kr and Xe in {the atmospheres of the} terrestrial planets and primitive CI meteorites. {The values shown for these gases are presented} relative to {their} solar abundances, in units of atoms per 10$^6$ Si atoms (adapted from Fig. 2 of Pepin 1991). {\bf The vertical arrow pointing down indicates that the Venus atmospheric abundance of Xenon is only an upper limit (see text and Donahue et al. 1981).}}
\label{fig1}
\end{figure}

\clearpage

\begin{table*}
\centering \caption{The abundances of heavy noble gases in terrestrial planet atmospheres in g/g-Planet (from Pepin 1991 and references therein).}
\begin{center}
\begin{tabular}{lcccc}
\hline
\hline
Planets      		& $^{20}$Ne						& $^{36}$Ar						& $^{84}$Kr						& $^{130}$Xe								\\
\hline
Venus     			& $2.9 \pm 1.3 \times 10^{-10}$		& $2.51 \pm 0.97 \times 10^{-9}$		& $4.7 ^{+0.6}_{-3.4} \times 10^{-12}$	& $8.9 ^{+2.5}_{-6.8} \times 10^{-14}$	       		\\              
Earth       			& $1.00 \pm 0.01 \times 10^{-11}$		& $3.45 \pm 0.01 \times 10^{-11}$		& $1.66 \pm 0.02 \times 10^{-12}$		& $1.40 \pm 0.02 \times 10^{-14}$	 	         		\\
Mars       			& $4.38 \pm 0.74 \times 10^{-14}$		& $2.16 \pm 0.55 \times 10^{-13}$		& $1.76 \pm 0.28 \times 10^{-14}$		& $2.08 \pm 0.41  \times 10^{-16}$		         	 	\\
\hline
\end{tabular}
\end{center}
\label{NG}
\end{table*}

\clearpage

\begin{table}
\centering \caption{{The ratio of observed noble gas abundances between the} terrestrial planets.}
\begin{center}
\begin{tabular}{lcc}
\hline
\hline
Noble gas       	& $X_{Venus}/X_{Earth}$		& $X_{Earth}/X_{Mars}$          	\\
\hline
$^{20}$Ne   	& 15.8--42.4                        		& 193.3--277.5                 		\\
$^{36}$Ar       	& 44.5--101.2                       	& 126.9--214.9                  		\\
$^{84}$Kr       	& 0.8--3.2                          		& 80.4--113.5                   		\\
$^{130}$Xe  	& 1.5--8.3                          		& 55.4--85.0                    		\\
\hline
\end{tabular}
\end{center}
\label{ratio}
\end{table}

\clearpage

 \begin{table}
\centering \caption{The ratio of {the number of objects impacting on Venus and Earth as a function of the number hitting the Earth and Mars, respectively} for each of the populations of comets used (from Horner et al. 2009).}
\begin{center}
\begin{tabular}{lcc}
\hline
\hline
Population      		& $N_{Venus}/N_{Earth}$		& $N_{Earth}$/$N_{Mars}$	\\
\hline
$N_{com-init}$     	& 0.86                          		& 3.57                      			\\
$N_{JNS}$       		& 0.78                          		& 2.94                      			\\
$N_{SNN}$       	& 0.82                          		& 3.70                      			\\
\hline
\end{tabular}
\end{center}
\label{ratio1}
\end{table}

\clearpage

\begin{table}
\centering \caption{The {mass fraction noble gas abundances of Venus and Earth as a function of those for Earth and Mars, as derived from Eq. 1.}}
\begin{center}
\begin{tabular}{lcccc}
\hline
\hline
Population      		& $X_{Venus}/X_{Earth}$         & $X_{Earth}/X_{Mars}$ 	\\
\hline
$N_{com-init}$      	& 1.05                     	 		& 0.38                  		\\
$N_{JNS}$       		& 0.95                      			& 0.32                  		\\
$N_{SNN}$       	& 1                         			& 0.40                  		\\
\hline
\end{tabular}
\end{center}
\label{ratio2}
\end{table}

\end{document}